\documentclass[12pt,a4paper,twoside]{article}

\usepackage{fancyhdr,graphicx,xcolor,young,ytableau}

\usepackage{amsmath,amsfonts,amsthm}
\usepackage[numbers,sort&compress,merge]{natbib}
\usepackage{authblk}
\usepackage[title,titletoc,toc]{appendix}

\DeclareMathOperator{\tr}{Tr}

\DeclareMathOperator{\Sc}{Sc}

\newcommand{\abs}[1]{\left \lvert #1 \right \rvert}
\newcommand{\av}[1]{\mathbb{E}\left\{ #1 \right \}} 

\theoremstyle{plain}

\theoremstyle{definition}

\begin{document}
\title{Moments of the eigenvalue densities and of the secular coefficients of $\beta$-ensembles\footnote{With the appendix \emph{A multivariate
Hermite polynomial identity} by Brian Winn, Department of Mathematical Sciences, Loughborough University, Loughborough LE11 3TU, UK}} 
\author{Francesco Mezzadri and Alexi~K. Reynolds\\
School of Mathematics, University of Bristol, Bristol BS8 1TW,
  UK}
\date{6 November 2016}
\maketitle

\abstract{We compute explicit formulae for the moments of the densities of the eigenvalues of the classical $\beta$-ensembles 
for finite matrix dimension as well as the expectation values of the coefficients of the characteristic polynomials. 
In particular, the moments are linear combinations of averages of Jack polynomials, whose coefficients are related to 
specific examples of Jack characters.}


\section{Introduction}
The density of the eigenvalues is of particular importance in the
study of random matrices for its intrinsic theoretical interest as
well as its many applications to various areas of physics.  One of the
main reasons is that the fluctuations of the eigenvalues around the
limiting density manifest on a global scale of the spectrum
in the properties of their linear statistics (see,
e.g.,~\cite{Joh98,DE06,DP12}), which play a primary role whenever a
mathematical or physical problem requires a probabilistic analysis.

Recently there has been a surge of interest in computing the
expectation value 
\begin{equation}
  \label{eq:moments}
    M_k =  \av{\tr X^k }_{\mathcal{E}}, \qquad k \in \mathbb{N},
\end{equation}
where $X$ is a $N \times N$ Hermitian matrix belonging to an
appropriate ensemble $\mathcal{E}$.  Eq.~\eqref{eq:moments} can also
be interpreted as the moments of the eigenvalue density.  There is an
extensive literature on the average~\eqref{eq:moments} as well as on
its large $N$ asymptotics when $\mathcal{E}$ is one of the 
Gaussian, Laguerre or Jacobi ensembles of real, complex or
quaternion matrices, usually labelled by
$\beta=1,2,4$. Indeed, The moments of the density of the eigenvalues of the Gaussian Unitary
Ensemble (GUE) have a particular important combinatorial meaning:
they count certain graphs embedded on a surfaces of a given genus $g$.
The $2g$-th coefficient in the asymptotic expansion
of~\eqref{eq:moments} for large $N$ are the number of pairings of $2g$
vertices in a regular polygon.  This idea was pioneered by Br\'ezin at
al.~\cite{BIPZ78}, and has since played a prominent role in quantum
field theory (see, e.g.,~\cite{Bou11} and references therein). When
$\mathcal{E}$ is the Jacobi or Laguerre ensemble with $\beta=1,2,4$
~\eqref{eq:moments} many important properties of the
electrical conductance and the Wigner delay time in chaotic quantum
cavities can be extracted from the
averages~\eqref{eq:moments}~\cite{Nov08,VV08,MS11,MS12,LV11,CMSV16a,CMSV16b}. 

In this article we shall derive explicit formulae for the moments of
the spectral densities of the Gaussian, Laguerre and Jacobi
$\beta$-ensembles as well as the averages of the secular coefficients of the characteristic polynomials.
The joint probability density function (j.p.d.f.)  of the eigenvalues is defined by
\begin{equation}
  \label{eq:jpdf}
  p_{\beta,w}(x_1,\dotsc,x_N) = \frac{1}{\mathcal{N}_{\beta}}
\prod_{j=1}^Nw(x_j) \prod_{1 \le j < k\le
    N}\abs{x_k - x_j}^\beta, 
\end{equation}
where $\beta$ can be any strictly positive real number and  
 \begin{equation}
\label{weights}
w(x) = \left \{ \begin{alignedat}{4}
& e^{-\frac{x^{2}}{2}},  &  x & \in \mathbb{R},  &  & &  &
   \text{Gaussian $\beta$-ensemble,} \\
 & x^{\gamma}e^{- x}, & x & \in \mathbb{R_+},  && \gamma > -1,
&&    \text{Laguerre $\beta$-ensemble,}  \\
&  x^{\gamma_1}(1-x)^{\gamma_2}, \quad  &  x & \in [0,1], &\quad &
\gamma_1,\gamma_2 > -1,
  &\quad & \text{Jacobi $\beta$-ensemble.}
   \end{alignedat} \right .
\end{equation}
The normalization constants $\mathcal{N}_{\beta}$ can be evaluated explicitly
using Selberg's integral. (For more details see, e.g.,~\cite[Sec.~4.7]{For10}.)

The $\beta$-ensembles were introduced by Dumitriu and Edelman~\cite{DE02}, and by Dumitriu~\cite{Dum03}, 
who developed tri-diagonal matrix models for the Gaussian and Laguerre
$\beta$-ensembles; Lippert~\cite{Lip03}, and Killip and Nenciu~\cite{KN04} discovered 
a sparse matrix model for the Jacobi $\beta$-ensemble.  Dumitriu et al.~\cite{DES07} designed a computer
algorithm that calculates the Jack polynomials and their averages in terms of other standard symmetric polynomials;  this software
can be used to compute  the moments~\eqref{eq:moments} iteratively. Recent 
articles~\cite{BG13a,BG13b,CE06,MMPS12,WF14,WF15,For11,CEM11,BEMPF12} have studied  the large $N$ expansions of the resolvent of $\beta$-ensembles using the loop equations formalism; in turn, this technique leads to  large $N$ expansion of the moments~\cite{WF14,WF15}.  Cunden et al.~\cite{CMV15} proved a formula for the covariances of the moments of one-cut 
$\beta$-ensembles in the limit $N\to \infty$. Fyodorov et al.~\cite{FDR10} used the moments of the Gaussian $\beta$-ensemble to compute a phase transition in the distribution of the velocities of a one-dimensional turbulent fluid satisfying the Burger equation.  Interestingly, in a recent paper Fyodorov and Le Doussal~\cite{FD16} showed that the moments of the  Jacobi $\beta$-ensemble  play a central role in the theory of the maximum of the GUE characteristic polynomials and log-correlated Gaussian processes, which recently have received much attention. 

In this article we will prove that the averages~\eqref{eq:moments} can be expressed as finite sums in terms of averages of Jack polynomials $C_\lambda^{(\alpha)}(x)$, $x=x_1,\dotsc,x_N$, namely
\begin{equation}
\label{main_result}
M_k = \frac{1}{\alpha^k k!}\, \sum_{\lambda \, \vdash k} j_\lambda\kappa^{\lambda}_{(k)}(\alpha) \, \mathbb{E}\left \{C_{\lambda}^{(\alpha)} \right \}_{\mathcal{E}}, \qquad k \le N,
\end{equation}
where $j_\lambda$ is a combinatorial factor which will be defined in Sec.~\ref{jackpol}, Eq.~\eqref{j_const}.
The expectation values in~\eqref{main_result} are with respect one of the \emph{j.p.d.f.}'s~\eqref{weights} with  the identification $\alpha=2/\beta$\footnote{This notation is not very common in the Random Matrix Theory literature, but it is more convenient in the theory of symmetric functions.  Because of the extensive use of Jack polynomials, we shall adopt the parametrization $\alpha=2/\beta$ throughout the paper.} and were computed by Kadell~\cite{Kad97}, and Baker and Forrester~\cite{BF97}; one of the main contributions of this paper is the derivation of  explicit formulae for~\eqref{main_result}.
The sum in~\eqref{main_result} is over all the partitions of  $\lambda = (\lambda_1,\cdots,\lambda_m)$ of $k$, 
denoted by$\lambda \vdash k$. The \emph{parts} $\lambda_j$ are integers such that 
\[
\lambda_1 \ge \dotsb \ge \lambda_m \ge 0, \qquad \abs{\lambda} = \lambda_1 + \dotsb + \lambda_m = k.
\] 

The formalism of the theory of symmetric function is very powerful in studying $\beta$-ensembles. As a corollary of the results on the moments, we compute the \emph{secular coefficients} of the expectation value of the characteristic polynomial.  More precisely, consider
\begin{equation}
P_{X}(z) = \det\left(X -z I\right) =(-1)^N \sum_{j=0}^N (-1)^j \Sc_j(X) z^{N-j}. 
\end{equation}
We show that 
\begin{equation}
\label{average_in} 
\mathbb{E}\left \{ \Sc_k(X)\right \}_{\mathcal{E}} = \frac{(\alpha)_k}{\alpha^k k!}\, \mathbb{E}\left\{ C_{(1^k)}^{(\alpha)}\right \}_{\mathcal{E}},
\end{equation}
where $(x)_k= \Gamma(x +k)/\Gamma(x)$ is the Pochhammer symbol. Haake et al.~\cite{HKSSZ96} and later Diaconis and Gamburd~\cite{DG04} computed all the joint moments of the \emph{secular coefficients} of the characteristic polynomial of a Haar distributed random unitary matrix.  It is far from obvious at this stage how to determine all the joint moments of the traces and of the secular coefficients for $\beta$-ensembles, as it seems beyond techniques available at present; we discuss
the reasons in Sec.~\ref{sec:jac_char} and in detail in Sec.~\ref{sec:example}.    

The structure of this paper is the following:  Sec.~\ref{jackpol} contains the definition and basic properties of the Jack polynomials; in Sec.~\ref{sec:jac_char} we introduce the Jack characters; in Sec.~\ref{inv_char} we compute the coefficients of the expansion~\eqref{main_result}; Sec.~\ref{mom_beta} gives the explicit formulae of the moments for the Laguerre, Jacobi and Gaussian $\beta$-ensembles; in Sec.~\ref{sec:neg_moment} we detail how to compute the negative moments; Sec.~\ref{sec:example} present an alternative derivation of the Jack characters, which illustrates the challenges one encounters when computing the 
joint moments; in Sec.~\ref{sec_coef} we compute the secular coefficients of the average of the characteristic polynomials; Sec.~\ref{conclusions} ends the paper with concluding remarks and an outlook on future research.

\section{Partitions and Jack Polynomials}
\label{jackpol}

Before discussing our results we need to introduce some notions from the theory of symmetric functions.  
\subsection{Partitions and Ferrers Diagrams}
\label{ssec:partition}
Let $\lambda$ be a partition of $k$  and let $\ell(\lambda)$ be \emph{the length} of $\lambda$, i.e. the number of parts of $\lambda$ different from zero.
It is sometimes convenient to represent $\lambda$ with a \emph{Ferrers diagram,} which is a table of $k$ boxes arranged 
in $\ell(\lambda)$ left-justified rows: the first row contains $\lambda_1$ boxes, the second $\lambda_2$, and so on.  We write $\lambda \subseteq \mu$ if the Ferrers diagram of $\lambda$ is contained in that of $\mu$, i.e. if $0\le \lambda_i \le \mu_i$.    
\begin{gather*}  
\begin{ytableau}
*(gray)  & *(gray)  & *(gray) & *(gray)&\\
*(gray) &*(gray) &&\\
*(gray) & \\
\\
\end{ytableau}\\
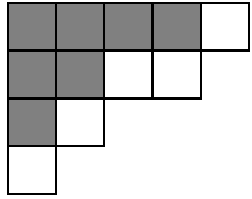
\text{\small{Fig. 1. The relation $(4,2,1) \subseteq (5,4,2,1)$ represented in terms of Ferrers diagrams.}}
\end{gather*}
Formally a Ferrers diagram of $\lambda$ can be defined as the set of points $(i,j) \in \mathbb{Z}^2$ such that $0\le j \le \lambda_i$. We write $ s \in \lambda$ if the box   $s=(i,j)$ belongs to the diagram of $\lambda$. The \emph{conjugate} $\lambda'$ of $\lambda$ is the partition whose Ferrers diagram is the transpose of $\lambda$, i.e. $\lambda' = (\lambda_1',\dotsc,\lambda_{m'}')$ such that
\begin{equation}
\lambda_i' = \# \left\{ j : \lambda_j \ge i \right \}.
\end{equation}
Note that $\lambda_1'=\ell(\lambda)$ and $\lambda_1=\ell(\lambda')$.  Given the Ferrers diagram of $\lambda$, we define
\begin{subequations}
\begin{align}
a_\lambda(s)& = a_\lambda(i,j)= \lambda_i -j  && \text{arm length,}\\
l_\lambda(s) & = \lambda'_j -i && \text{leg length,}\\
h^*_\lambda(s) & = l_\lambda(s) + \alpha(1 +a_\lambda(s))  && \text{upper hook length,}\\
h_*^\lambda(s) & = l_\lambda(s) + 1 +\alpha a_\lambda(s)  && \text{lower hook length.}
\end{align}
\end{subequations}
The arm length is the number of boxes to the right of $s=(i,j)$; the leg length is the number of boxes below $s$. Similarly, the co-arm length $a'_\lambda(s) = j-1$ and co-leg length $l'_\lambda(s)=i-1$ are the number of boxes to the left of and above $s$, respectively.   The parameter $\alpha >0$ is the same that defines the scalar product~\eqref{scalar_product}.
 \begin{gather*}
 \begin{ytableau}
*(white)  &   &  &  & & \\
  & s &*(gray) & *(gray) & *(gray)\\
 & *(gray) & \\
& *(gray) \\
\end{ytableau}
\qquad \qquad \qquad
 \begin{ytableau}
*(white)  &   &  &  & & \\
  &*(gray) s &*(gray) & *(gray) & *(gray)\\
 & *(gray) & \\
& *(gray) \\
\end{ytableau}\\
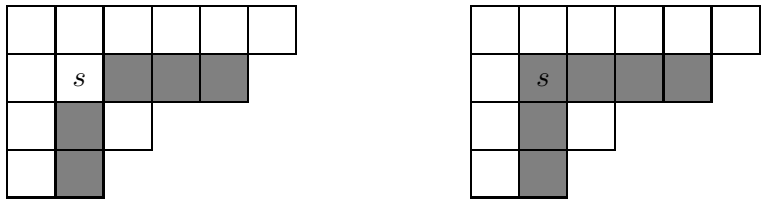
\text{\small{Fig.~2. The shaded areas (left diagram) are the arm length $a_\lambda(s)=3$ and leg}}\; \\
\text{\small{length $l_\lambda(s)=2$ of $s$ for $\lambda=(6,5,3,2)$. The shaded area in the right diagram}} \\
\text{\small{is the hook length $h_*^\lambda(s)=h^*_\lambda(s)=6$ for $\alpha=1$.}} \qquad \qquad \qquad \qquad 
\qquad \qquad \;\;\;\; {}
 \end{gather*}
It is often convenient to define the following quantities:
\begin{subequations}
\begin{align}
\label{c_const}
c'(\lambda,\alpha) &= \prod_{s \in \lambda} h_\lambda^*(s) \\
\label{cp_const}
c(\lambda,\alpha) &=\prod_{s\in \lambda}h_*^\lambda(s)\\
\label{j_const}
j_\lambda & = c(\lambda,\alpha)c'(\lambda,\alpha)
\end{align}
\end{subequations} 
In this paper we will often use the multivariate generalization of the Pochhammer symbol,
\begin{equation}
\label{gpochhammer}
\begin{split}
\left(t\right)^\alpha_\lambda & = \prod_{i=1}^{\ell(\lambda)} \left(t - \frac{i-1}{\alpha}\right)_{\lambda_i} = \prod_{i=1}^{\ell(\lambda)} \frac{\Gamma\left(t - \frac{i-1}{\alpha} + \lambda_i\right)}{\Gamma\left(t - \frac{i-1}{\alpha} \right)}\\
& = \prod_{i=1}^{\ell(\lambda)} \prod_{j=1}^{\lambda_i}\left(t - \frac{i-1}{\alpha} + j - 1\right)
=  \prod_{s \in \lambda} \left(t - \frac{l'_\lambda(s)}{\alpha} + a'_\lambda(s)\right), \qquad \alpha > 0.
\end{split}
\end{equation}

We introduce a total ordering in the set of partitions of an integer $k$ by saying that $\lambda > \mu$  whenever $\lambda_i - \mu_i$ is strictly positive for the first index $i$ such that $\lambda_i\neq \mu_i$.   This is known as \emph{lexicographical} ordering.  For example,
\begin{equation*}
(3,3,2,1) > (3,3,1,1,1).
\end{equation*} 
If $\lambda > \mu$ we say that the monomial $x_1^{\lambda_1} \dotsb x_N^{\lambda_N}$ 
is of \emph{higher weight} than $x_1^{\mu_1} \dotsb x_N^{\mu_N}$. Another ordering on the set of partitions 
of particular relevance to the the theory of the Jack polynomials is the \emph{dominance,}
or \emph{natural,} ordering.  We say that $\mu$ ``dominates'' $\lambda$, $\lambda \preceq \mu$, if 
\begin{equation}
\label{dominance_ord}
\lambda_1 + \dotsb + \lambda_i \le \mu_1 + \dots + \mu_i, \qquad 1\le i < \max\left\{\ell(\lambda),\ell(\mu)\right\}.
\end{equation}
If any of the inequalities is strict we write $\lambda \prec \mu$.  It is worth emphasizing that both the lexicographical and dominance orderings compare partitions of the same integer.  The dominance ordering is a partial ordering as soon as $\abs{\lambda}=\abs{\mu} \ge 6$; for example, the partitions $(4,1,1)$ and $(3,3)$ cannot be compared.  If $\lambda \preceq \mu$ then $\lambda \le \mu$, but the opposite is not necessarily true.

\subsection{Jack Polynomials}

In the following we shall adopt the notation $x=(x_1,\dotsc,x_N)$, unless it is evident from the context that $x\in \mathbb{R}$. Let $\lambda \vdash k$ and $\ell(\lambda) = m \le N$, the Jack polynomial $C_\lambda^{(\alpha)}(x)$ is a symmetric, homogeneous polynomial that satisfies the following properties:
\begin{itemize}
\item[(i)]  The term of highest weight in $C_\lambda^{(\alpha)}(x)$ is $x_1^{\lambda_1}x_2^{\lambda_2} \dotsb x_m^{\lambda_m}$, that is
\begin{equation}
C_\lambda^{(\alpha)} (x)= d_\lambda  x_1^{\lambda_1}x_2^{\lambda_2} \dotsb x_m^{\lambda_m} +  \, \text{terms of lower weight,}
\end{equation}
where $d_\lambda$ is a constant.
\item[(ii)]  $C_\lambda^{(\alpha)}$ is an eigenfunction of the differential operator
\begin{equation}
\label{caloger-sutherland}
\Delta^{(\alpha)}  = \sum_{j=1}^N  x_j^2 \frac{\partial^2}{\partial x^2_j}  + 
\frac{2}{\alpha}\sum_{\substack{j,k=1\\ j \neq k}}^N \frac{x_j^2}{x_j - x_k} \frac{\partial}{\partial x_j}.
\end{equation}
\item[(iii)]  The normalization of $C_\lambda^{(\alpha)}$  is fixed by the condition
\begin{equation}
\label{normalization}
(x_1 + \dotsb + x_N)^k = \sum_{\substack{\lambda \, \vdash k\\ \ell(\lambda) \le N}}C_\lambda^{(\alpha)}(x).
\end{equation}
 \end{itemize}

It can be shown that the statements (i)-(iii) define the Jack polynomials uniquely.  A nice proof is presented in Muirhead~\cite[Sec. 7.2.1]{Mui82} for the Zonal polynomials ($\alpha=2$), but it can be easily generalized to any $\alpha >0$.   The operator $\Delta^{(\alpha)}$ is (up to a similarity transformation) the Hamiltonian of a Calogero-Sutherland quantum many-body system (see~\cite{BF97} for the details).  The polynomials $C_\lambda^{(\alpha)}$ are non-degenerate eigenfunctions of $\Delta^{(\alpha)}$ with eigenvalues $\rho_\lambda^\alpha + k(N-1)$, where
\begin{equation}
\label{eig}
\rho_\lambda^\alpha= \sum_{j=1}^N\lambda_j \left(\lambda_j - 1 -\frac{2}{\alpha}\left(j-1\right)\right).
\end{equation} 
A formula that will be useful in the rest of the paper is
\begin{equation}
\label{C_id}
C^{(\alpha)}_\lambda \left(1^N\right) = \frac{\alpha^{2\abs{\lambda}}\abs{\lambda}!}{j_\lambda}\left(\frac{N}{\alpha}\right)_{\lambda}^\alpha,
\end{equation}
where $1^N=(1,\dotsc,1)$ (see, e.g.,~\cite[Sec.~2]{DES07}).

There are equivalent definitions of the Jack polynomials that lead to different normalizations (see, e.g.,~\cite[Sec.~VI.10]{Mac95} and~\cite[Sec.~2]{DES07}).    These differences are summarised nicely in Dumitriu et al.~\cite[Sec.~2]{DES07}.    The `C' definition that we adopt is more natural for studying $\beta$-ensembles and Selberg-type integrals, as it appears in the theory of the scalar hypergeometric functions of matrix argument.   The other two common definitions are the `P' and `J' normalizations.   In the `P' definition the coefficients of the monomial of highest weight is required to be one;  the `J' normalization sets the coefficient of the monomial $x_1\dotsb x_k$ (known as the \emph{trailing} coefficient) to $k!$, where $\abs{\lambda}=k$.  Their relations to the `C' definition are  
\begin{subequations}
\begin{align} 
\label{Preno} 
P^{(\alpha)}_\lambda (x) &= \frac{c'(\lambda,\alpha)}{\alpha^{\abs{\lambda}}\abs{\lambda}!} C_\lambda^{(\alpha)}(x),\\
\label{J_norm}
J^{(\alpha)}_{\lambda}(x) & = \frac{j_{\lambda}}{\alpha^{\abs{\lambda}} \abs{\lambda}!} \, C^{(\alpha)}_\lambda (x).
\end{align}
\end{subequations}

A closed formula for the Jack polynomials does not exist, but they can be computed using certain recurrence relations involving the \emph{monomial symmetric functions}
\begin{equation}
\label{monomial}
m_\lambda(x) = \sum_{\sigma \in \mathfrak{S}_N} x_{\sigma 1}^{\lambda_1}x_{\sigma 2}^{\lambda_2} \dotsb x_{\sigma N}^{\lambda_N}.
\end{equation}
It turns out that
\begin{equation}
\label{P_jack_rec}
P^{(\alpha)}_{\lambda}= m_{\lambda} + \sum\limits_{\sigma \prec \lambda} u_{\lambda \sigma}^{\alpha} \, m_{\sigma}.
\end{equation}
The coefficients $u^\alpha_{\lambda \sigma}$ can be calculated recursively and 
can be used to construct the Jack polynomials explicitly from~\eqref{P_jack_rec}~\cite{Mac95,DES07}. 

Traces of matrices are particular cases of \emph{power sum symmetric functions,}
\begin{equation}
p_\lambda =p_{\lambda_1}\dotsb p_{\lambda_m}, \qquad p_{\lambda_j}= x_1^{\lambda_j} + \dotsb + x_N^{\lambda_j}.
\end{equation}
Denote by $\Lambda^k_N$ the ring of homogeneous symmetric  polynomials of degree $k$ in $N$ variables. The Jack polynomials and the power sum symmetric functions form two sets of bases in $\Lambda_N^k$. We can define the scalar product
\begin{equation}
\label{scalar_product}
\left \langle p_\lambda,p_\mu  \right \rangle_\alpha = \alpha^{\ell(\lambda)} z_\lambda \delta_{\lambda 
\mu},
\end{equation}
where $z_\lambda = 1^{r_1} r_1!\, 2^{r_2}r_2!\, \dotsb k^{r_k} r_k!$ and $r_j$ denotes the number of 
times $j$ appears in the partition $\lambda$.  The power sum symmetric functions play a prominent role in the theory of the Jack polynomials due to the following orthogonality relation
\begin{equation}
\label{orthogonality}
\left \langle P^{(\alpha)}_\lambda ,P_\mu^{(\alpha)}\right \rangle_{\alpha} = 0 \quad \text{unless $\lambda = \mu.$}
\end{equation}
One can show that Eqs.~\eqref{P_jack_rec} and~\eqref{orthogonality} define the multivariate polynomials $P^{(\alpha)}_\lambda$ uniquely~\cite[Sec. VI.10]{Mac95}.   We also have  (see Stanley~\cite[Th.~5.8]{Sta89})
\begin{equation}
\label{norm}
\left \langle J_\lambda^{(\alpha)}, J_\mu^{(\alpha)} \right \rangle_{\alpha}= j_\lambda \delta_{\lambda \mu}.
\end{equation}

%
%
%
%
%

\section{Jack characters}
\label{sec:jac_char}

Since the Jack polynomials and the power sum symmetric functions both form bases in $\Lambda_N^k$, they are related by a linear transformation.  The coefficients that express the Jack polynomials in terms of the power sum symmetric functions are known as \emph{Jack characters} and play an important role in combinatorics.  Using standard notation we write
\begin{equation}
\label{j_charJ}
J^{(\alpha)}_\lambda  = \sum_{\mu \, \vdash k}\theta_\mu^\lambda(\alpha) p_\mu. 
\end{equation}
In this paper we are interested in the inverse transformation, namely
\begin{equation}
\label{inv_jcharJ}
p_\mu  = \sum_{\lambda \, \vdash k} \kappa_\mu^{\lambda} (\alpha) J_\lambda^{(\alpha)}.
\end{equation} 
 Writing these maps in the `C' normalization gives
 \begin{subequations}
 \begin{align}
\label{jack_characters}
C_\lambda^{(\alpha)} & = \frac{\alpha^{k} k!}{j_\lambda}\, \sum_{\mu \, \vdash k} \theta_\mu^{\lambda} (\alpha) p_\mu \\
\label{invthetag}
p_\mu &= \frac{1}{\alpha^{k} k!}\, \sum_{ \lambda \, \vdash k} j_\lambda 
 \kappa^{\lambda}_{\mu}(\alpha) \, C_{\lambda}^{(\alpha)}.
\end{align}
\end{subequations}

In the context of  Random Matrix Theory (RMT) the averages of the $p_\mu$'s are the joint moments
\begin{equation}
\label{joint}
\mathbb{E} \left\{\prod_{j=1}^k \left(\tr X^j\right)^{r_j}\right\}_{\mathcal{E}}, \qquad 1 \leq r_1 + 2r_2 + \dotsb + kr_k \le N.
\end{equation}
The RHS of Eq.~\eqref{invthetag} depends on the eigenvalues only through the Jack polynomials $C_\lambda^\alpha$.  Therefore, in order to compute their expectation values
we need to determine the coefficients $\kappa_\mu^\lambda(\alpha)$, which are independent of 
the ensemble.  The knowledge of the coefficients $\kappa_\mu^\lambda(\alpha)$ is equivalent to that of 
the Jack characters $\theta_\mu^\lambda(\alpha)$, which at present is beyond reach in its full generality.  Nevertheless, we can
ask if we can compute few particular cases.  The first can be obtained trivially from the normalization
 condition~\eqref{normalization} and formula~\eqref{invthetag}, which give
\begin{equation}
\label{kp_1k}
\kappa^{\lambda}_{(1^k)}(\alpha) = \frac{\alpha^k k!}{j_\lambda}.
\end{equation} 
In this paper we will compute two other particular cases.  We will find an explicit
formula for $\kappa_{(k)}^\lambda(\alpha)$,\footnote{The subscript $(k)$ denotes a partition of length $1$ and size $k$.} which allows us to determine the expectation values of 
\begin{equation}
\label{invthetachar}
p_k = \frac{1}{\alpha^k k!}\, \sum_{ \lambda \, \vdash k} j_\lambda \kappa^{\lambda}_{(k)}(\alpha) \, C_{\lambda}^{(\alpha)}.
\end{equation}
In addition, in Sec.~\ref{sec:example} we will be able to evaluate the average of
\begin{equation}
\label{part_case}
 p_{\mu}= \tr X^2 \left(\tr X \right)^{k - 2}, 
\end{equation}
where
\begin{equation}
\mu = (2,1^{k - 2}) = (2,\underbrace{1,\dotsc,1}_{k - 2 \; \text{times}}).
\end{equation}
Unfortunately, the methods applied to these two examples do not seem to extend to the general case.

In order to understand the difficulties involved in computing~\eqref{joint} (or equivalently $\theta^\lambda_\mu(\alpha)$), consider the next most complicated case from~\eqref{invthetachar},  the correlations $\mathbb{E}\left \{ \tr X^j \tr X^k \right \}$. One may observe that by simply multiplying the RHS of Eq.~\eqref{invthetachar} we obtain a sum involving the products $C_\lambda^{(\alpha)} C_\mu^{(\alpha)}$.  Then, there are two possible approaches.  The product $C_\lambda^{(\alpha)} C_\mu^{(\alpha)}$ is a homogeneous symmetric polynomial, which can be written in the basis of Jack polynomials; if we can compute the coefficients of this linear combination, we can then average the resulting sum using the formulae in Sec.~\ref{sec:gauss_mom}.   Alternatively, we could attempt to average $C_\lambda^{(\alpha)} C_\mu^{(\alpha)}$ directly.  In both cases we would have to compute the coefficients of the linear combination
\begin{equation}
\label{HLcoef}
C_\mu^{(\alpha)}C_\nu^{(\alpha)} = \sum_{\lambda} c_{\mu \nu}^\lambda C^{(\alpha)}_\lambda,
\end{equation}
or  equivalently the scalar product $\left \langle C_\lambda^{(\alpha)}, C_\mu^{(\alpha)}C_\nu^{(\alpha)} \right \rangle$.  In the special case of the Schur functions, i.e. $\alpha=1$, the $c^\lambda_{\mu \nu}$ reduce to the Littlewood-Richardson coefficients.
It turns out that this problem  is equivalent to that of computing the $\theta^\mu_\lambda(\alpha)$'s (see, e.g.,
 Stanley~\cite{Sta89}).  We discuss the technical challenges encountered in this problem in detail in Sec.~\ref{sec:example}.

The study of the Jack characters was initiated by Hanlon~\cite{Han88}, who conjectured a first combinatorial interpretation;
Stanley~\cite{Sta89} proved various properties, among which he derived explicit formulae for few specific cases. Recently there has been a surge of interest in the Jack characters in the combinatorics literature~\cite{Las08,Las09,FS11,DFS14,Vas13,DF13,DF14,KV14}, as they play a central role in in the theory of symmetric functions. In particular, Do\l\c{e}ga and F\'{e}ray~\cite{DF14} showed that they are polynomials in $\alpha$ with rational coefficients; Kanunnikov and Vassilieva~\cite{KV14} proved a recurrence relation for them. At present a general expression for arbitrary partitions $\lambda, \mu$ and any $\alpha >0$ is still lacking. 

When $\alpha=1$ ($\beta=2$) the Jack polynomials reduce to the Schur functions $s_\lambda$; more precisely we have $P^{(1)}_\lambda = s_\lambda$. Then, Eq.~\eqref{jack_characters} becomes
\begin{equation}
\label{schur_fun}
s_\lambda = \sum_{\mu\, \vdash m} z_\mu^{-1}\chi_{\mu}^\lambda \, p_\mu,
\end{equation} 
where are the $\chi^\lambda_\mu$'s are the characters of the irreducible representations of the symmetric group $\mathfrak{S}_m$ (see, e.g.,~\cite[Sec. I.7]{Mac95}).\footnote{The irreducible representation of $\mathfrak{S}_m$ are labelled by the partitions of $m$.    The notation $\chi^\lambda_\mu$ denotes the character of the irreducible representation $\lambda$ evaluated on permutations of cycle-type $\mu$. } This leads to 
\begin{equation}
\theta_\mu^\lambda(1)=\frac{c(\lambda,1)}{z_\mu}\chi_{\mu}^\lambda.
\end{equation}
The inverse relation of~\eqref{schur_fun}  is 
\begin{equation}
\label{invshur}
p_\mu = \sum_{\lambda \, \vdash m} \chi_{\mu}^\lambda \, s_\lambda. 
\end{equation}
When $\alpha=2$ the Jack polynomials reduce to the \emph{Zonal polynamials} and the $\theta_\mu^\lambda(2)$ are known as \emph{Zonal characters.} F\'{e}ray and \'{S}niady~\cite{FS11} expressed the Zonal characters as sums over pair-partitions. 

A consequence of the invariance of the j.p.d.f.'s \eqref{eq:jpdf} under permutations of the eigenvalues is that the theory of symmetric functions appears naturally in many areas of RMT, whenever an algebraic or combinatorial structure of the ensemble can be exploited by the formalism of symmetric functions.  For example, formulae~\eqref{schur_fun} and~\eqref{invshur}, together with the fact that the Schur functions are the characters of the unitary group, were used by Diaconis and Shahshahani~\cite{DS94} to prove that the joint moments of traces of Haar distributed unitary matrices are those of independent standard complex random variables.  Notwithstanding their name,  when $\alpha \neq 1$ the Jack characters are not associated to any group. However, their averages with respect to the j.p.d.f.'s~\eqref{weights} are known~\cite{BF97}; these formulae combined with the theory of the Jack polynomials and of special functions of matrix argument have allowed us to compute explicit formulae for the coefficients $\kappa_{(k)}^\lambda(\alpha)$ and hence for the moments~\eqref{eq:moments}.    

\section{The coefficients $\kappa_{(k)}^\lambda(\alpha)$}
\label{inv_char}

In order to compute the moments~\eqref{eq:moments} in closed form we need two ingredients: the explicit expression of the coefficients of the expansion~\eqref{invthetachar} and the averages $\mathbb{E}\left\{ C^{(\alpha)}_\lambda \right\}$.   The generalized hypergeometric functions admit expansions in terms of Jack polynomials (see, e.g.~\cite[Ch. 13]{For10}) and in particular cases they can be turned into generating functions of  certain symmetric functions.  This allows the explicit evaluation of the coefficients $\kappa_{(k)}^\lambda(\alpha)$ as well as the expansion of the elementary symmetric functions $e_i$ of the eigenvalues.   Here we show that
\begin{equation}
\label{f_inv_jc}
\kappa_{(k)}^\lambda (\alpha) = \frac{\alpha k}{j_\lambda} \prod_{s \in \lambda \setminus \{(1,1)\}}
\left(\alpha a'_\lambda (s) - l'_\lambda(s)\right), \qquad \abs{\lambda}=k.
\end{equation}
Writing in terms of the coordinates in the Ferrers diagram $s=(i,j)$ and using
$a'_\lambda(s)= j-1$ and $l_\lambda(s)=i-1$ gives the equivalent formula
\begin{equation}
\label{f_inv_jc2}
\kappa_{(k)}^\lambda (\alpha) =(-1)^{\abs{\lambda}- \lambda_1} \frac{\alpha^k k (\lambda_1 - 1)!}{j_\lambda} 
\prod_{i=1}^{\ell(\lambda)-1}\binom{i/\alpha}{\lambda_{i+1}}\lambda_{i+1}!.
\end{equation}

Consider the generating function
\begin{equation}
\label{p_gen_fun}
\begin{split}
P(t)& =\sum_{r \ge 1}p_r t^{r-1}  =  \sum_{i=1}^N \sum_{r\ge 1} x_i^r t^{r-1}\\
&= \sum_{i=1}^N \frac{x_i}{1-x_i t}  = \sum_{i = 1}^N \frac{d}{dt} \log\frac{1}{1-x_i t}.
\end{split}
\end{equation}
These series are formal and there is no statement of convergence about them.  Integrating both sides gives
\begin{equation}
\label{int_gen}
\sum_{r\ge 1}\frac{p_r}{r}t^r =- \sum_{i=1}^N \log\left(1 - x_i t \right).
\end{equation}
We now write the RHS as
\begin{equation}
\label{log_gen}
\sum_{i= 1}^N \log\left(1 - x_i t \right)= 
\lim_{u \to 0} \frac{d}{du} \prod_{i= 1}^N\left(1 - x_i t\right)^u
\end{equation}
and use a generalization of the binomial theorem (see Forrester~\cite[Pro. 13.1.11]{For10})
\begin{equation}
\label{gen_bin}
 \,_1\mathcal{F}_0^{\left( \alpha \right)} \left( a; t^N;x_1,\dotsc,x_N \right) =\prod\limits_{i=1}^N \left(1 - x_i t\right)^{-a},
\end{equation}
where $t^N$ denotes the $N$-component vector $(t,\dotsc,t)$ and $a$ is a real parameter.  The LHS denotes a hypergeometric function of two sets of variables, which can be expressed in terms of the series 
\begin{equation}
\label{F_01}
 \,_1\mathcal{F}_0^{\left( \alpha \right)} \left(a; x_1,\dotsc,x_N ;y_1,\dotsc,y_N\right) 
= \sum\limits_{\lambda} \frac{\left(a \right)^{ \alpha}_\lambda}{| \lambda |!}  \frac{C^{\left( \alpha \right)}_\lambda \left(x_1,\dotsc,x_N\right) C^{\left( \alpha \right)}_\lambda \left(y_1,\dotsc,y_N \right)}{C^{\left( \alpha \right)}_\lambda \left(1^N \right) }.
\end{equation}
This is a particular case of the more general hypergeometric function
\begin{multline}
\label{hyper_geo_two_v}
{}_{p}\,\mathcal{F}^{(\alpha)}_q(a_1,\dotsc,a_p;b_1,\dotsc,b_q;x_1,\dotsc,x_N;y_1,\dotsc,y_N)\\
= \sum_{\lambda} \frac{1}{\abs{\lambda}!} \frac{(a_1)^{\alpha}_\lambda \dotsb \, (a_p)^{\alpha}_\lambda}{(b_1)^{\alpha}_\lambda \dotsb \, (b_q)^{\alpha}_\lambda} \frac{C^{(\alpha)}_\lambda(x_1,\dotsc,x_N)C^{(\alpha)}_\lambda(y_1,\dotsc,y_N)}{C^{(\alpha)}_\lambda \left(1^N\right)}.
\end{multline}
Since the Jack polynomials are homogeneous $C_\lambda^{(\alpha)} \left(t^N\right)= t^{\abs{\lambda}} C_\lambda^{(\alpha)} \left( 1^N\right)$, which combined with~\eqref{F_01} and~\eqref{log_gen} gives
\begin{equation}
\label{lim_u}
\lim_{u \to 0} \frac{d}{du} \prod_{i= 1}^N\left(1 - x_i t\right)^u
=   \sum\limits_{\lambda} \frac{1}{|\lambda|!} t^{|\lambda|} \left( \lim_{u \rightarrow 0} \frac{d}{du}  \left( -u \right)^{ \alpha }_\lambda \right) C^{\left( \alpha \right)}_\lambda \left(x\right).
\end{equation}
Comparing the coefficients of the powers of $t$ in ~\eqref{int_gen} and~\eqref{lim_u}  leads to
\begin{equation}
\label{kappa_lim}
\kappa^{\lambda}_{(k)}(\alpha) =-\frac{\alpha^k k}{j_\lambda} \lim_{u \to 0}
\frac{d}{du}\left(-u\right)^{\alpha}_\lambda.
\end{equation}

The generalised Pochhammer's symbol $\left(-u\right)^{\alpha}_\lambda$ is a polynomial in $u$, whose coefficient of the linear term gives the limit~\eqref{kappa_lim}.  Finally, from Eq.~\eqref{gpochhammer} the fact that for any partition $l'_\lambda((1,1))= a'_\lambda((1,1))=0$, we obtain 
\begin{equation}
\begin{split}
\lim_{u\to 0}\frac{d}{du} \left(-u\right)^{\alpha}_\lambda & = \lim_{u\to 0} \frac{(-u)}{u} \prod_{s \in \lambda \setminus \{(1,1)\}}\left(-u -\frac{l'_\lambda (s)}{\alpha} + a'_\lambda (s) \right)\\
& = - \alpha^{1-k} \prod_{s \in \lambda \setminus \{(1,1)\}}\left(\alpha a'_\lambda (s) -l'_\lambda (s)\right),
\end{split}
\end{equation} 
which inserted into~\eqref{kappa_lim} gives~\eqref{f_inv_jc}.

In the combinatorics literature, the objects of interest are the coefficients $\theta_\mu^\lambda (\alpha)$ in the expansion~\eqref{C_id}, rather than their inverse $\kappa_\mu^{\lambda}(\alpha)$.  The two quantities, however, are connected by a simple relation of proportionality. This follows from the observation that both the $p_\lambda$'s and $J_\lambda^{(\alpha)}$'s are orthogonal with respect to the scalar product~\eqref{scalar_product} and
\begin{equation*}
\left \langle J_\lambda ^{(\alpha)}, J_\mu^{(\alpha)} \right \rangle_\alpha 
= \sum_\rho \theta_\rho ^{\lambda}(\alpha) \theta_\rho^{\mu}(\alpha) \alpha^{\ell(\rho)} z_\rho = \delta_{\lambda \mu}\, j_\mu.
\end{equation*}
By definition the $\kappa_\lambda^\mu(\alpha)$ is the inverse of the transfer matrix of $\theta_\lambda^\mu (\alpha)$, therefore
\begin{equation}
\label{inv_th}
\theta_\lambda^\mu (\alpha) = \frac{j_\mu}{\alpha^{\ell(\lambda)} z_\lambda} \kappa_\lambda^\mu (\alpha).
\end{equation}
When $\lambda =(k)$, $\ell(\lambda)=1$ and $z_\lambda = k$; therefore,  Eq.~\eqref{f_inv_jc} gives 
\begin{equation}
\label{mac_ex}
\theta_{(k)}^\lambda (\alpha) =  \prod_{s \in \lambda \setminus \{(1,1)\}}\left(\alpha a'_\lambda (s) - l'_\lambda (s)\right).
\end{equation}
This formula for $\theta_{(k)}^\lambda (\alpha)$ can be found without proof in the book by Macdonald~\cite[Ch. VI.10, Example 1(b)]{Mac95}. An alternative expression for $\theta_{(k)}^\lambda (\alpha)$ can also be found in~\cite{Sta89} and in the book by Forrester --- equation (12.145) of \cite{For10} is the equivalent of \eqref{invthetachar} with the coefficients $\kappa_{(k)}^\lambda(\alpha)$ explicitly given by quantities defined in earlier sections of that reference.

\section{The Moments of $\beta$-Ensembles}
\label{mom_beta}

In Sec.~\ref{inv_char} we computed the coefficients of the expansions of the symmetric polynomials $p_k$ in terms of the Jack symmetric functions.   In order to compute the moments of the eigenvalue densities we need the expectation values of the Jack polynomials, which were obtained by Kadell~\cite[Th.~1]{Kad97} for the Jacobi $\beta$-ensemble and by Baker and Forrester~\cite[Cor.~3.2 and Cor.~4.1]{BF97} for the Gaussian and Laguerre $\beta$-ensembles. Averages with respect to the measures of the $\beta$-ensembles are evaluated by introducing multivariate generalizations of the classical Hermite, Laguerre and Jacobi polynomials. We briefly summarise their basic definitions and properties in Appendices~\ref{app:A}, \ref{app:B} and \ref{app:C}.


\subsection{Laguerre $\beta$-ensemble}
\label{lag_ssec}

The classical Laguerre polynomials can be generalized to multivariate homogeneous polynomials that are 
orthogonal with respect to the measure of the Laguerre $\beta$-ensemble.  The theories of the Jack and of the 
multivariate Laguerre polynomials are intertwined, since  the Laguerre polynomials can 
be expressed as linear combinations of the Jack polynomials (see Eq.~\eqref{mult_L_exp}, Appendix~\ref{app:A}). It turns out that 
\begin{equation}
\label{av_jack}
\begin{split}
\mathbb{E}\left \{ C^{(\alpha)}_\lambda \right \}_{\textrm{L}\beta\textrm{E}} &  = \frac{1}{\mathcal{N}^{(\mathrm{L})}_\alpha} \int_{[0,\infty)^N} 
C_\lambda^{(\alpha)}(x)\prod_{j=1}^N x^\gamma_j e^{-x_j}
\prod_{1 \le j < k \le N}\abs{x_k - x_j}^{2/\alpha}d^N x \\
& = \bigl(\gamma + 1 +(N-1)/\alpha \bigr)^{\alpha}_\lambda C_\lambda^{(\alpha)} \left(1^N\right)= L_{\lambda}^{\alpha, \gamma}(0), 
\end{split}
\end{equation} 
where $\mathcal{N}_\alpha^{(\mathrm{L})}$ is the normalization constant of the Laguerre ensemble, $0$ is the origin in $\mathbb{R}^N$ and $d^Nx=dx_1\dotsb dx_N$.  When $N=1$ this average reduces to the classical formula
\begin{equation*}
L_{k}^{\gamma}\left(0 \right) = \frac{1}{\Gamma\left( \gamma+1 \right)} \int_0^{\infty}x^k x^\gamma e^{-x} dx.
\end{equation*}

Finally Eq.~\eqref{av_jack} gives 
\begin{equation}
\label{mom_L_expl}
\begin{split}
M_k^{(\textrm{L})}& = \frac{1}{\alpha^k k!}\, \sum_{\lambda \, \vdash k} j_\lambda \kappa^{\lambda}_{(k)}(\alpha) 
\, L_{\lambda}^{\alpha, \gamma}(0)\\
 & =  \frac{1}{\alpha^k k!}\, \sum_{\lambda \, \vdash k} j_\lambda \kappa^{\lambda}_{(k)}(\alpha) \, 
\bigl(\gamma + 1 +(N-1)/\alpha \bigr)^{\alpha}_\lambda C_\lambda^{(\alpha)} \left(1^N\right), \qquad k \le N.
\end{split}
\end{equation}
where the coefficients $\kappa^{\lambda}_{(k)}(\alpha)$ are given in Eq.~\eqref{f_inv_jc} and $C_\lambda^{(\alpha)} \left(1^N\right)$ in Eq.~\eqref{C_id}.  This is one of the main results of this paper.  Formula~\eqref{mom_L_expl} gives an explicit and self-contained expression for the moments of the L$\beta$E.

\subsection{Jacobi $\beta$-ensemble}
\label{ssec:jacobi}

The Jacobi polynomials have a multivariate generalization too~\cite{Kad97,BF97,DES07}. 
The formula analogous  to~\eqref{av_jack} is
\begin{equation}
\label{av_jack_J}
\begin{split}
\mathbb{E}\left \{ C^{(\alpha)}_\lambda \right \}_{\textrm{J}\beta\textrm{E}}  & = 
\frac{1}{\mathcal{N}^{(\mathrm{J})}_\alpha} \int_{[0,1]^N} 
C_\lambda^{(\alpha)}(x)\prod_{j=1}^N x_j^{\gamma_1} (1-x_j)^{\gamma_2} 
\prod_{1 \le j < k \le N}\abs{x_k - x_j}^{2/\alpha}d^N x \\
&= \frac{\bigl( \gamma_1 + (N-1)/\alpha +1 \bigr)_\lambda^{ \alpha }}{\bigl(\gamma_1 + \gamma_2 + 2(N-1)/\alpha +2 \bigr)_\lambda^{\alpha }} \, C_\lambda^{(\alpha)} \left(1^N \right)=J_{\lambda}^{\alpha, \gamma_1, \gamma_2}(0),
\end{split}
\end{equation} 
which for $N=1$ becomes
\begin{equation}
J_{k}^{\gamma_1, \gamma_2}\left(0 \right) = \frac{\Gamma\left(2+\gamma_1 +\gamma_2\right)}{\Gamma\left( \gamma_1+1 \right)\Gamma\left( \gamma_2+1 \right)} \int_0^1x^k x^{\gamma_1} \left(1- x\right)^{\gamma_2} dx.
\end{equation} 
Finally, we arrive at 
\begin{equation}
\label{mom_J_expl}
\begin{split}
M_k^{(\textrm{J})} & =  \frac{1}{\alpha^k k!}\, \sum_{\lambda \, \vdash k} 
j_\lambda \kappa^{\lambda}_{(k)}(\alpha) \, J_{\lambda}^{\alpha, \gamma_1, \gamma_2}(0)\\
& =  \frac{1}{\alpha^k k!}\, \sum_{\lambda \, \vdash k} j_\lambda \kappa^{\lambda}_{(k)}(\alpha) \, 
\frac{\bigl( \gamma_1 + (N-1)/\alpha +1 \bigr)_\lambda^{ \alpha }}{\bigl(\gamma_1 + \gamma_2 + 2(N-1)/\alpha +2 \bigr)_\lambda^{\alpha }} \, C_\lambda^{(\alpha)} \left(1^N\right), \qquad k \le N.
\end{split}
\end{equation}

\subsection{Gaussian $\beta$-ensemble}
\label{sec:gauss_mom}

The calculation of the moments of the Gaussian $\beta$-ensemble follows a similar pattern.  When $\abs{\lambda}$ is even we have  
\begin{equation}
\label{av_jack_Hermite}
\begin{split}
\mathbb{E}\left \{ C^{(\alpha)}_\lambda \right \}_{\textrm{G}\beta\textrm{E}} &  = 
\frac{1}{\mathcal{N}^{(\mathrm{G})}_\alpha} \int_{\mathbb{R}^N} 
C_\lambda^{(\alpha)}(x)\prod_{j=1}^N e^{-\frac{x_j^2}{2}}
\prod_{1 \le j < k \le N}\abs{x_k - x_j}^{2/\alpha}d^N x \\
& = (-1)^{\frac{\abs{\lambda}}{2}} H_\lambda^{\alpha}(0),  
\end{split}
\end{equation} 
where $H_\lambda^{\alpha}(x)$ are the multivariate Hermite polynomials (see Appendix~\ref{app:C}).  When $\abs{\lambda}$ is odd the symmetry of the integrand implies that the average~\eqref{av_jack_Hermite} is zero.  When $N=1$ the expectation values can be written as   
\begin{equation}
H_k(0)=   \frac{(-1)^{\frac{k}{2}}}{\sqrt{2\pi}}\int_{-\infty}^\infty x^k e^{-\frac{x^2}{2}} dx,
\end{equation}
where the $H_k(x)$'s are the classical Hermite polynomials. The moments are
\begin{equation}
\label{moments_hermite}
M_k^{(\textrm{H})}  =  \frac{(-1)^{\frac{k}{2}}}{\alpha^k k!}\, \sum_{\lambda \, \vdash k} 
j_\lambda \kappa^{\lambda}_{(k)}(\alpha) \, H_{\lambda}^{\alpha}(0).
\end{equation}
To our knowledge an explicit formula for $H_\lambda^{\alpha}(0)$ does not exist; however, in Appendix~\ref{brian_proof} we present a proof by Brian Winn of a particular case, namly
\begin{equation}
\label{H_conj}
H^{\alpha}_{(1^k)}(0) = 
\begin{cases}
\frac{\alpha^{k/2}k!(k-1)!!}{(\alpha)_k}\binom{N}{k} & \text{if $k$ is even,}\\
0 & \text{if $k$ is odd.}
\end{cases}
\end{equation}
This formula will become useful in the next section.  For general partition $H_\lambda^{\alpha}(0)$ can be evaluated 
using the Maple routine MOPS~\cite{DES07}, which computes Jack polynomials and multivariate Laguerre, 
Jacobi and Hermite polynomials symbolically.

\section{Negative moments of $\beta$-ensembles}
\label{sec:neg_moment}

The formulae discussed in Sec.~\ref{mom_beta} can be used to compute the negative moments of the Laguerre and Jacobi $\beta$-ensembles. It was brought to our attention during the writing of this paper that it is also possible to calculate some
explicit formulae for the negative integer moments of the Laguerre and Jacobi $\beta$-ensembles. Formulae for negative moments of
the Jacobi $\beta$-ensemble first appeared in~\cite{FD16}. We outline here a simple derivation for calculating negative integer
moments via Jack polynomial theory. Denote by $1/x$ the vector $(1/x_1,\dotsc,1/x_N)$ and consider~\eqref{invthetachar} in which $x \mapsto 1/x$,
\begin{equation}
\label{neg1}
p_k\left(1/x\right) = \frac{1}{\alpha^k k!}\, \sum_{ \lambda \, \vdash k} j_\lambda \kappa^{\lambda}_{(k)}(\alpha) \, C_{\lambda}^{(\alpha)}\left(1/x\right).
\end{equation}
There exists an interesting functional relation between $C_{\lambda}^{(\alpha)}(x)$ and $C^{(\alpha)}_\lambda(1/x)$, namely
\begin{equation}
\label{forrel2}
C_\lambda^{(\alpha)} \left(1/x\right) = 
\frac{k!\, \alpha^{2k-Nt} }{\left(Nt-k\right)!} \frac{c'\left(\left(t^N-\lambda\right)^{+}, \alpha \right)}{ c'\left(\lambda,\alpha\right)} x_1^{-t} \dotsb x_N^{-t} C_{\left(t^N-\lambda\right)^{+}}^{(\alpha)} \left(x\right),
\end{equation}
where $N \ge \ell(\lambda)$, $t \ge \lambda_1$ is an integer and 
\begin{equation}
\left(t^N-\lambda\right)^{+} = \left(t, \dots, t, t-\lambda_{\ell\left(\lambda\right)}, \dots, t-\lambda_1\right).
\end{equation}
Equation~\eqref{forrel2} is the expression in the `C'-normalization of a formula that can be found in~\cite[p. 643]{For10}. 

The negative moments can be computed by substituting~\eqref{forrel2} into~\eqref{neg1} and calculating the corresponding expectation values.  
\begin{itemize}
\item Laguerre $\beta$-ensemble: 
\begin{equation}
\begin{split}
\mathbb{E}\left \{ x_1^{-t} \dotsc x_N^{-t} C_{\left(t^N-\lambda\right)^{+}}^{(\alpha)} \left(x\right) \right \}_{\textrm{L}\beta\textrm{E}} &= \mathbb{E}\left \{ C_{\left(t^N-\lambda\right)^{+}}^{(\alpha)} \left(x\right) \right \}_{\textrm{L}\beta\textrm{E}}\bigg|_{\gamma \rightarrow \gamma - t} \\
 & = L_{\left(t^N - \lambda\right)^+}^{\alpha,\gamma - t}(0).
\end{split}
\end{equation}
The parameter $\gamma$ is the exponent in the integrand~\eqref{av_jack}; clearly this average exists only if $\gamma - t >-1$, i.e. $\gamma > \lambda_1 - 1$.
\item Jacobi beta ensemble:
\begin{equation}
\begin{split}
\mathbb{E}\left \{ x_1^{-t} \dotsb x_N^{-t} C_{\left(t^N-\lambda\right)^{+}}^{(\alpha)} \left(x\right) \right \}_{\textrm{J}\beta\textrm{E}} & = \mathbb{E}\left \{ C_{\left(t^N-\lambda\right)^{+}}^{(\alpha)} \left(x\right) \right \}_{\textrm{J}\beta\textrm{E}}\bigg|_{\gamma_1 \rightarrow \gamma_1 - t}\\
&  = J_{(t^N - \lambda)^+}^{\alpha,\gamma_1 - t,\gamma_2}(0),
\end{split}
\end{equation}
provided that $\gamma_1-t > -1$ (i.e. $\gamma_1 > \lambda_1 - 1$).
\end{itemize}
In~\cite{FD16} the authors find that the explicit formulae for negative moments of the Jacobi $\beta$-ensemble do not depend on the choice of $t$ and that $t$ may be set to zero to find a simple final expression.  

\section{Higher Order Correlations:  Discussion and an Example} 
\label{sec:example}
In Sec.~\ref{sec:jac_char} we argued that the computation of the joint moments~\eqref{joint} is beyond our present ability as it involves the knowledge of the complete set of the Jack characters $\theta^\mu_\lambda(\alpha)$.  Here we detail a unified proof of three particular examples, which combined with the results in Sec.~\ref{mom_beta}, give the expectation value~\eqref{eq:moments} and 
\begin{equation}
\mathbb{E}\left\{\left(\tr X\right)^k \right\}_{\mathcal{E}}, \qquad  
\mathbb{E}\left\lbrace\tr X^2 \left(\tr X \right)^{k - 2} \right \rbrace_{\mathcal{E}}. 
\end{equation}This proof is particularly instructive because it gives clear evidence that computing 
the Jack characters in full generality is beyond the techniques available at present.  

Consider the map $\epsilon_Y$ defined on the ring $\Lambda_N^k$ by 
\begin{equation}
\epsilon_Y\left(p_\lambda\right) = Y^{\ell(\lambda)},
\end{equation} 
where $Y$ is an arbitrary parameter.   One can show that~\cite[Ch. VI.10, Eq.~(10.25)]{Mac95} 
\begin{equation}
\label{macmap}
\epsilon_Y\left(J_\lambda^{(\alpha)}\right) =  \prod_{s \in \lambda} \left(Y + a'_\lambda(s)\alpha - l'_\lambda(s)\right) = \frac{\left(Y/\alpha\right)^\alpha_{\lambda}}{\alpha^{\abs{\lambda}}} ,   
\end{equation}
where $a'_\lambda(s)$ and $l'_\lambda(s)$ are the co-arm and co-leg lengths of $s\in \lambda$, and $(t)^\alpha_\lambda$
is the generalised Pochhammer symbol (see Sec.~\ref{ssec:partition}). Substituting Eq.~\eqref{j_charJ} into the LHS of~\eqref{macmap} gives
\begin{equation}
\sum_{\mu \, \vdash k}\theta_\mu^\lambda(\alpha) Y^{\ell(\mu)} = \prod_{s \in \lambda} \left(Y + a'_\lambda(s)\alpha - l'_\lambda(s)\right). 
\end{equation} 
This equation implies 
\begin{equation}
\label{mac_for}
\sum_{\substack{\mu \, \vdash k\\ \ell(\mu) = j}} \theta_\mu^\lambda(\alpha)  = (-1)^j e_{k - j},
\end{equation}
where the elementary symmetric functions $e_j$ are evaluated at $l'_\lambda(s)-a'_{\lambda}(s) \alpha$ 
as $s$ varies in the Ferrers diagram of $\lambda$.   When there exists only one partition $\mu\vdash k$ such that $\ell(\mu) =j$, Eq.~\eqref{mac_for} gives a formula for the Jack character $\theta_\mu^\lambda(\alpha)$.  For arbitrary integers $k$ there are
at least three partitions with this property:
\begin{subequations}
\begin{align}
\label{f_par}
\mu & = (k), & \ell(\mu) &=1, \\
\label{s_par}
\mu &= (1^k) = (\underbrace{1,\dotsc,1}_{k \; \text{times}}), & \ell(\mu) & = k,\\
\label{t_par}
\mu &= (2,1^{k-2}) = (2,\underbrace{1,\dotsc,1}_{k -2 \; \text{times}}), &  \ell(\mu) &= k-1.
\end{align}
\end{subequations}  
Equation~\eqref{mac_for} combined with~\eqref{f_par} gives formula~\eqref{mac_ex} (note that $a_\lambda'(1,1)= l'_\lambda(1,1)=0$);  the second case gives $\theta_{(1^k)}^\lambda(\alpha) =1$, which is consistent with Eqs.~\eqref{kp_1k} and~\eqref{inv_th};  the partition leads to 
\begin{equation}
\theta^\lambda_{(2,1^{k-2})}(\alpha) = \alpha \sum_{j =1}^{\ell(\lambda')} (j - 1)\lambda'_j - \sum_{i=1}^{\ell(\lambda)} (i-1)\lambda_i.
\end{equation}
This formula can also be found in~\cite[Ch. VI.10, p. 348]{Mac95}.

The derivation in its section starts from formula~\eqref{macmap}, which is quite subtle and whose proof is far from trivial.   It is based on Pieri formula and on the duality
\begin{equation}
\omega_\alpha\left(P^{(\alpha)}_\lambda \right) = b_{\lambda'}^{(\alpha^{-1})} P_{\lambda'}^{(\alpha^{-1})},  
\end{equation}
where 
\begin{equation}
b_\lambda^{(\alpha)} = \prod_{s \in \lambda} \frac{\alpha a_\lambda(s) + l_\lambda(s) + 1}{\alpha a_\lambda(s) + l_\lambda(s) + \alpha}.
\end{equation}
Pieri formula gives the product $P_\lambda^{(\alpha)} e_\mu $ as a linear combination of $P_\lambda^{(\alpha)}$, whose coefficients can be explicitly computed in terms of certain combinatorial expressions.  It can be thought as first step toward calculating the generalization of the Littlewood-Richardson  coefficients~\eqref{HLcoef}.  Other Pieri-type formulae are available, but they are far from leading to an expression for the Littlewood-Richardson coefficients.  This in turns means that computing the average~\eqref{joint} or even the expectation values $\mathbb{E}\left \{ \tr X^j \tr X^k \right \}$ is beyond the techniques available at
the moment.

\section{The Secular Coefficients}
\label{sec_coef}

Consider 
\begin{equation}
P_{X}(z) = \det\left(X -z I\right) =(-1)^N \sum_{j=0}^N (-1)^j \Sc_j(X) z^{N-j}. 
\end{equation}
The quantities $\Sc_k(X)$ are the \emph{secular coefficients.}  Note that
\[
\Sc_0(X)=1, \qquad \Sc_1(X)= \tr (X), \qquad \Sc_N(X) = \det(X). 
\]
The secular coefficients of $P_X(z)$ are symmetric polynomials of the eigenvalues.  More precisely, 
\begin{subequations}
\begin{align}
e_k(x_1,\dotsc,x_N) & = \Sc_k(X),
\intertext{where the quantities} 
e_k(x_1,\dotsc,x_N) & = \sum_{i_1 < \dotsb < \, i_k} x_{i_1} \dotsb \, x_{i_j}
\end{align}
\end{subequations}
are known as \emph{elementary symmetric polynomials.}  For any partition $\lambda$ 
we define
\begin{equation}
e_\lambda (x_1,\dotsb,x_N) = e_{\lambda_1}\dotsb e_{\lambda_m}.
\end{equation}
The elementary symmetric functions form a basis in the ring of symmetric polynomials of a given degree and as such can be expressed in terms of the Schur function, namely
\begin{equation}
\label{esym_funct}
e_\mu = \sum_{\lambda \, \vdash k}K_{\lambda' \mu}  s_\lambda.
\end{equation}
The entries of the transition matrix $K_{\lambda \mu}$ are \emph{the Kostka numbers,} which give the number of semi-standard Young tableaux of shape $\lambda$ and weight $\mu$.\footnote{A semi-standard Young tableaux is a Ferrers diagram filled with integers that are weakly increasing along the rows and strictly decreasing down the columns; the partition formed by the number of times a given integer appears in the tableaux is the weight.}  

The first ones to address the problem of computing the averages of the secular coefficients of random unitary matrices were Haake et al.~\cite{HKSSZ96}. Diaconis and Gamburd~\cite{DG04} used~\eqref{esym_funct} and the fact that the Schur functions are the characters of the irreducible representation of the unitary group to compute all the joint moments of $\Sc_k(X)$.    We may ask what is the generalization of~\eqref{esym_funct} in terms of Jack polynomials; the answer to this question would allow us to solve the analogous problem for characteristic polynomials of $\beta$-ensembles.  As for the Jack characters at the moment we can only give a partial solution.    We first notice that by definition $e_k = m_{(1^k)}$ and that $(1^k) \prec \lambda $ for any $\lambda \vdash k$; therefore, by~\eqref{P_jack_rec} 
$e_k = P^{(\alpha)}_{(1^k)}$ and by~\eqref{Preno}
\begin{equation}
\label{eq:4}
e_k = \frac{c'\bigl( (1^k),\alpha\bigr)}{\alpha^k k!}C_{(1^k)}^{(\alpha)} .
\end{equation}
Now, $a(s)=0$ for any $s \in (1^k)$; therefore,
\begin{equation}
c'\bigl((1^k),\alpha\bigr) = \prod_{s\in (1^k)}\left(l_{(1^k)}(s) + \alpha \right)=\prod_{j=0}^{k-1} \left(j + \alpha \right) = \left(\alpha \right)_{k}.
\end{equation}
Finally, we arrive at 
\begin{equation}
\label{average} 
\mathbb{E}\left \{ \Sc_k(X)\right \}_{\mathcal{E}} = \frac{(\alpha)_k}{\alpha^k k!}\, \mathbb{E}\left\{ C_{(1^k)}^{(\alpha)}\right \}_{\mathcal{E}}.
\end{equation}

The explicit expressions for each $\beta$-ensemble for the averages~\eqref{average} can be computed using the expectation values~\eqref{av_jack}, \eqref{av_jack_J}, \eqref{av_jack_Hermite} and the formula
\begin{equation}
\label{eq:5}
C_{(1^k)}^{(\alpha)}\left(1^N\right) = \frac{\alpha^k k!}{(\alpha)_k}\binom{N}{k}.
\end{equation} 
We obtain
\begin{subequations}
\begin{align}
\mathbb{E}\left \{ \Sc_k(X)\right \}_{\mathrm{L}\beta\mathrm{E}} & = 
\frac{(\alpha)_k}{\alpha^k k!}\, L^{\alpha, \gamma}_{(1^k)}(0)  \nonumber \\
& = \left(\gamma + 1 +\frac{N-1}{\alpha} \right)^{\alpha}_{(1^k)} \binom{N}{k}, \\
\mathbb{E}\left \{ \Sc_k(X)\right \}_{\mathrm{J}\beta\mathrm{E}} &=\frac{(\alpha)_k}{\alpha^k k!}\, 
J^{\alpha, \gamma_1,\gamma_2}_{(1^k)}(0) \nonumber \\
&= \frac{\bigl( \gamma_1 + (N-1)/\alpha +1 \bigr)_{(1^k)}^{\alpha}}%
{\bigl(\gamma_1 + \gamma_2 + 2(N-1)/\alpha +2 \bigr)_{(1^k)}^{\alpha }} \binom{N}{k}, \\
\mathbb{E}\left \{ \Sc_k(X)\right \}_{\mathrm{G}\beta\mathrm{E}} &= 
(-1)^{\frac{k}{2}}\frac{(\alpha)_k}{\alpha^k k!}\, H^{\alpha}_{(1^k)}(0) \nonumber \\
\label{gaussian_sec}
&= \begin{cases}
\left(-1\right)^{k/2} 
\frac{(k-1)!!}{\alpha^{k/2}}\binom{N}{k} & \text{if $k$ is even,} \\
0 & \text{if $k$ is odd.}
\end{cases}
\end{align}
\end{subequations}
where in the last line we have used formula~\eqref{H_conj}.



\section{Conclusions}
\label{conclusions}

We computed the positive and negative moments of the density of the eigenvalues 
and the averages of the  secular coefficients for the Gaussian, 
Laguerre and Jacobi $\beta$-ensembles for matrices of finite dimensions.  Our approach is based on 
the theory of the Jack polynomials, which are a natural tool in the study of $\beta$-ensembles.   The Jack polynomials form a 
basis in the ring of homogeneous symmetric  polynomials.  As such they can be expressed in terms of other symmetric 
functions like the power sum, the monomial and the elementary symmetric functions.  The coefficients that express 
the Jack polynomials in terms of power sum symmetric functions are known as \emph{Jack characters}, which recently 
has been object of intense study in the combinatorics  literature~\cite{DF14,DFS14,FS11,KV14,Kad97,Las08,Las09,Vas13}.
Surprisingly, however, little is known about them and an explicit formula is not available.   
Since the traces of powers of matrices are particular cases of power sum symmetric functions, they can be expressed as linear
combinations of Jack polynomials, whose coefficients are a subset of the inverse of the Jack characters.   We were able 
to compute these coefficients explicitly and hence the moments of the density of states.  The expectation values of the secular
coefficients can also be expressed in terms of averages of Jack polynomials.

It is still an open question how to compute all the joint moments of the density of the eigenvalues of the $\beta$-ensembles.   Their knowledge is tantamount to having a complete understanding of the Jack characters, which at present is out of reach.  Similarly, evaluating all of the joint moments of the secular coefficients is equivalent to knowing the transition matrix from the Jack polynomials to the elementary symmetric functions;  the elements of this matrix are generalizations of the Kostka numbers.

\section*{Acknowledgements}

We are grateful to Pierre Le Doussal and Yan Fyodorov for bringing our attention toward formula~\eqref{forrel2} 
and to Brian Winn for writing Appendix D with the proof of Eq.~\eqref{H_conj}.  We also thank  
Marcel Novaes for helpful discussions.  FM was partially supported by EPSRC research grant EP/L010305/1.  
No empirical or experimental data was created during this study.

\begin{appendices}
\section{Multivariate Laguerre polynomials} 

\label{app:A}

Here we briefly introduce the generalized Laguerre polynomials and discuss the properties that were used in Sec.~\ref{lag_ssec}.  The exposition in this appendix follows the theory in~\cite{BF97,DES07}.

The classical Laguerre polynomials $L_k^{\gamma}(x)$, $k=0,1,\dotsc,$ are the unique (up to a constant) polynomials orthogonal in $[0,\infty)$ with respect to the weight $x^\gamma e^{-x}$, $\gamma >-1$. The generalized Laguerre polynomials $L_\lambda^{\alpha,\gamma}(x_1,\dotsc,x_N)$  are uniquely specified (up to a constant) as the polynomial part of the eigenfunctions of the Calogero-Sutherland operator
\begin{equation}
H^{(\mathrm{L})} = \sum_{j=1}^N x_j \frac{\partial^2}{\partial x_j^2}  +\left(\gamma - x_j + 1\right)\frac{\partial}{\partial x_j}
+\frac{2}{\alpha}\sum_{\substack{j,k=1\\ j \neq k}}^N \frac{x_j^2}{x_j - x_k} \frac{\partial}{\partial x_j}.
\end{equation}
 Now take the measure 
\begin{equation}
\label{lag_meas2}
d\mu^{(\mathrm{L})}(x) = \frac{1}{\mathcal{N}^{(\mathrm{L})}}\prod\limits_{i=1}^{N} x_{i}^{\gamma} e^{-x_i} \prod\limits_{1 \leq j < k \leq N} \abs{x_{k} - x_{j}}^{2/\alpha}d^Nx
\end{equation}
where the normalization constant $\mathcal{N}^{(\mathrm{L})}$ can be computed using Selberg's integral,
\begin{equation}
\label{norm_const}
\begin{split}
\mathcal{N}^{(\mathrm{L})}_{\alpha} & = \int_{[0,\infty)^N} \prod\limits_{i=1}^{N} x_{i}^{\gamma} e^{-x_i} \prod\limits_{1 \leq j < k \leq N} \abs{x_{k} - x_{j}}^{2/\alpha}d^Nx\\
& = \prod\limits_{i=0}^{N-1} \frac{\Gamma \left(1+(i+1)/\alpha\right) \Gamma \left(1 + i/\alpha+\gamma\right)}{\Gamma \left(1+1/\alpha\right)}.
\end{split}
\end{equation}
Let $f(x_1,\dotsc,x_N)$ and $g(x_1,\dotsc,x_N)$ homogeneous polynomials in $\mathbb{R}^N$ and define the scalar product 
\begin{equation}
\label{scalar_prod}
\left \langle f | g\right \rangle^{(\mathrm{L})}= \int_{[0,\infty)^N}f(x)g(x) d\mu(x).
\end{equation}
It turns out that the multivariate Laguerre polynomials are homogeneous and orthogonal with respect to the scalar product $\left \langle \ | \ \right \rangle^{(\mathrm{L})}$ and possess many features that generalize well known formulae for the classical Laguerre polynomials.

Since the generalized Laguerre polynomials are homogeneous, they can be represented as linear combination of Jack polynomials, 
\begin{equation}
\label{linear_comb_jack_L}
L_\lambda^{\alpha,\gamma}(x) = \sum_{\mu \subseteq \lambda }c_{\lambda \mu}^{\alpha} \, C_\mu^{(\alpha)}(x). 
\end{equation}
The coefficients $c_{\lambda \mu}^{\alpha}$ are uniquely determined by fixing the coefficient of highest weight; the normalization used in this paper is $c^{\alpha}_{\lambda \lambda}=(-1)^{\abs{\lambda}}$.  The explicit expansion of $L^{\alpha, \gamma}_\lambda (x)$ in terms of Jack polynomials is
\begin{equation}
\label{mult_L_exp}
L_{\kappa}^{\alpha, \gamma}(x) = \left( \gamma + \frac{N-1}{\alpha} +1 \right)_\lambda^{\alpha} C_\kappa^{\left(\alpha\right)} \left( 1^N \right) \sum\limits_{\sigma \subseteq \lambda} \frac{\left(-1\right)^{\abs{\sigma}} \binom{\lambda}{\sigma}}{\left( \gamma + \frac{N-1}{\alpha} +1 \right)_\sigma^{\alpha}}\frac{C_\sigma^{\left(\alpha\right)} \left( x \right)}{C_\sigma^{\left(\alpha\right)} \left( 1^N \right)},
\end{equation}
where the generalized binomial coefficients $\binom{\lambda}{\sigma}$  are defined by the expansion
\begin{equation}
\frac{C^{(\alpha)}_\lambda\left(1 + t_1,\dotsc,1 +t_N\right)}{C_\lambda^{(\alpha)}\left(1^N\right)} = \sum_{s=0}^{\abs{\lambda}} \sum_{\abs{\sigma}=s} \binom{\lambda}{\sigma}\frac{C_\sigma^{(\alpha)}(t)}{C_\sigma^{(\alpha)}\left(1^N\right)}.
\end{equation}
Since $\binom{\lambda}{(0)}=1$ setting $x=0$ in Eq.~\eqref{mult_L_exp} gives 
\begin{equation}
L_{\lambda}^{\alpha, \gamma}(0) =\left(\gamma + \frac{N-1}{\alpha} +  1\right)^{(\alpha)}_\lambda C_\lambda^{(\alpha)} \left(1^N\right).
\end{equation}
%

We now present the explicit expression of the first three moments calculated via~\eqref{mom_L_expl}:
\begin{subequations}
\label{thre_mom_L}
\begin{align}
M^{(\mathrm{L})}_1 &= N \left( \gamma+1- \frac{1}{\alpha} \right) + \frac {N^2}{\alpha},\\
M^{(\mathrm{L})}_2 &= N \left( \gamma^{2}+ 3\gamma - \frac{4}{\alpha} -\frac{3 \gamma}{\alpha}+2+ \frac{2}{\alpha^2} \right) + N^2 \left( \frac {3\gamma}{\alpha} - \frac{4}{\alpha^2}
+ \frac{4}{\alpha} \right) +\frac {2N^3}{\alpha^{2}},\\
M^{(\mathrm{L})}_3 
&= N \left(\gamma^3 - \frac{6}{\alpha^3} + 11\gamma + \frac{11\gamma}{\alpha^2} + \frac{17}{\alpha^2} - \frac{17}{\alpha} + 6 + 6\gamma^2 - \frac{6 \gamma^2}{\alpha} - \frac{21\gamma}{\alpha} \right) \nonumber \\
&\quad + N^2 \left(\frac{21\gamma}{\alpha} + \frac{17}{\alpha^3} + \frac{6 \gamma^2}{\alpha} + \frac{17}{\alpha} - \frac{21\gamma}{\alpha^2} - \frac{33}{\alpha^2} \right )\nonumber \\
&\quad + N^3 \left(\frac{16}{\alpha^2} - \frac{16}{\alpha^3} + \frac{10 \gamma}{\alpha^2} \right) +  \frac{5N^4}{\alpha^3}.
\end{align}
\end{subequations}

\section{Multivariate Jacobi polynomials} 
\label{app:B}

The theory of the multivariate Jacobi polynomials follows the same pattern~\cite{Kad97,BF97,DES07}.  They are identified up to a constant as the polynomial part of the eigenfunctions of the operator
\begin{equation}
\begin{split}
H^{(\mathrm{J})} & = \sum_{j=1}^N x_j(1-x_j) \frac{\partial^2}{\partial x_j^2}  +\bigl(\gamma_1 + 1 - x_j\left(\gamma_1 + \gamma_2 + 2\right)\bigr)\frac{\partial}{\partial x_j}\\
& \quad +\frac{2}{\alpha}\sum_{\substack{j,k=1\\ j \neq k}}^N \frac{x_j(1-x_j)}{x_j - x_k} \frac{\partial}{\partial x_j}.
\end{split}
\end{equation}
The multivariate Jacobi polynomials are orthogonal with respect to the measure 
\begin{equation}
\label{jac_meas}
d\mu^{(\mathrm{J})}(x) = \frac{1}{\mathcal{N}^{(\mathrm{J})}_\alpha}\prod\limits_{i=1}^{N} x_{i}^{\gamma_1} \left(1-x_j\right)^{\gamma_2} \prod\limits_{1 \leq j < k \leq N} \abs{x_{k} - x_{j}}^{2/\alpha}d^Nx, 
\end{equation}
where $x_j \in [0,1), \, \gamma_1,\gamma_2>-1$.  The normalization constant is 
\begin{equation}
\label{norm_constJ}
\begin{split}
\mathcal{N}^{(\mathrm{J})}_\alpha & = \int_{[0,1)^N} \prod\limits_{i=1}^{N} x_{i}^{\gamma_1} (1-x_j)^{\gamma_2} \prod\limits_{1 \leq j < k \leq N} \abs{x_{k} - x_{j}}^{2/\alpha}d^Nx\\
&  = \prod\limits_{i=0}^{N-1} \frac{\Gamma \bigl(1+\frac{i+1}{\alpha}\bigr) \Gamma \bigl(\gamma_1 + \frac{i}{\alpha}+1\bigr) \Gamma \bigl(\gamma_2 + \frac{i}{\alpha}+1\bigr)}{\Gamma \bigl(1+\frac{1}{\alpha}\bigr)\Gamma \bigl(\gamma_1 + \gamma_2 + \frac{N+i-1}{\alpha}+2\bigr)}.
\end{split}
\end{equation}
The Jacobi polynomial admit an expansion in terms of Jack polynomials too:
\begin{equation}
\begin{split}
J_{\lambda}^{\alpha, \gamma_1, \gamma_2}(x) & = \left( \gamma_1 + \frac{N-1}{\alpha} +1 \right)_\lambda^{\left( \alpha \right)} C_\kappa^{\left(\alpha\right)} \left( 1^N \right) \\
& \quad \times \sum\limits_{\sigma \subseteq \lambda} \frac{\left(-1\right)^{\abs{\sigma}} v_{\lambda \sigma}^\alpha}{\left( \gamma_1 + \frac{N-1}{\alpha} +1 \right)_\sigma^{\left( \alpha \right)}}\frac{C_\sigma^{\left(\alpha\right)} \left( x\right)}{C_\sigma^{\left(\alpha\right)} \left( 1^N \right)}.
\end{split}
\end{equation}
The coefficients $v_{\lambda \sigma}^\alpha $ obey the recurrence relation
\begin{equation}
\label{rec_v}
\begin{split}
v^\alpha_{\lambda \sigma}& = \frac{1}{\left(\bigl(\gamma_1 + \gamma_2 + \frac{2}{\alpha}\left(N-1\right)+2\bigr)\left(\abs{\lambda}-\abs{\sigma}\right) + \rho^{\alpha}_\lambda -\rho^{\alpha}_\sigma \right)} \\
& \quad \times \sum_{i \: \text{allowable}} \binom{\lambda}{\sigma^{(i)}}\binom{\sigma^{(i)}}{\sigma}v^\alpha_{\lambda \sigma^{(i)}}.
\end{split}
\end{equation}
where $\rho^{\alpha}_\lambda$ was defined in~\eqref{eig} and $\sigma^{(i)}=(\sigma_1,\dotsc,\sigma_i +1,\dotsc,\sigma_N)$.  It can be proved that the denominator in~\eqref{rec_v} never vanishes if $\sigma \subset \lambda$~\cite[Lemma 2.23]{DES07} and therefore the  recurrence relation is well defined.  The coefficients $v^{\alpha}_{\lambda \sigma}$ are uniquely defined by setting $v^{\alpha}_{\lambda \lambda}=1$ for all $\alpha$ and $\lambda$.  One can show that~\cite{DES07}
\begin{equation}
\begin{split}
J_{\lambda}^{\alpha, \gamma_1, \gamma_2}(0)= \frac{\bigl( \gamma_1 + (N-1)/\alpha +1 \bigr)_\lambda^{\left( \alpha \right)}}{\bigl(\gamma_1 + \gamma_2 + 2(N-1)/\alpha +2 \bigr)_\lambda^{\left( \alpha \right)}} \, C_\lambda^{\left(\alpha\right)} \left(1^N \right).
\end{split}
\end{equation}

%
%

We present here the first two moments calculated via formula~\eqref{mom_J_expl}:
\begin{subequations}
\begin{align}
M^{(\mathrm{J})}_1 & = \frac {N \left( \gamma_{1}\alpha+N-1+\alpha \right)}{\gamma_{2}\alpha+ \gamma_{
1}\alpha+2N-2+2\alpha},\\
M^{(\mathrm{J})}_2 
&= \frac{N}{\left( \gamma_{2}\alpha + \gamma_{1}\alpha + 2N - 3+2\alpha \right)} \frac{\left( \gamma_{1}\alpha + N - 1 + \alpha \right)}{\left( \gamma_{2}\alpha+ \gamma_{1}\alpha + 2N - 2 + 3\alpha \right)} \nonumber \\
& \quad \times  \frac{1}{\left( \gamma_{2}\alpha + \gamma_{1}\alpha + 2N - 2 + 2\alpha \right) } \left( \gamma_{1}^{2}\alpha^{2} + 4\gamma_{1}\alpha^{2} + \gamma_{1}\alpha^{2}\gamma_{2} + 4\alpha^{2} + 2\gamma_{2} \alpha^{2} \right.\nonumber \\
&\quad  \left.+ 7 N \alpha + 2N\gamma_{2}\alpha+3\gamma_{1}\alpha N - 9\alpha - 4\gamma_{1}\alpha - 2\gamma_{2}\alpha + 4 - 7N + 3N^{2} \right).
\end{align}
\end{subequations}

\section{Multivariate Hermite polynomials} 
\label{app:C}
The Calogero-Sutherland operator for the multivariate Hermite polynomials is
\begin{equation}
H^{(\mathrm{H})} = \sum_{j=1}^N  \frac{\partial^2}{\partial x_j^2}  -2x_j \frac{\partial}{\partial x_j}
+\frac{2}{\alpha}\sum_{\substack{j,k=1\\ j \neq k}}^N \frac{1}{x_j - x_k} \frac{\partial}{\partial x_j}.
\end{equation}
The measure with respect to which they are orthogonal is 
\begin{equation}
\label{her_meas}
d\mu^{(\mathrm{H})}(x) = \frac{1}{\mathcal{N}^{(\mathrm{H})}_\alpha}\prod\limits_{i=1}^{N} e^{-\frac{x_i^2}{2}} \prod\limits_{1 \leq j < k \leq N} \abs{x_{k} - x_{j}}^{2/\alpha}d^Nx,
\end{equation}
where 
\begin{equation}
\label{
H}
\begin{split}
\mathcal{N}^{(\mathrm{H})}_\alpha & = \int_{\mathbb{R}^N} \prod\limits_{i=1}^{N} e^{-\frac{x_i^2}{2}} \prod\limits_{1 \leq j < k \leq N} \abs{x_{k} - x_{j}}^{2/\alpha}d^Nx\\
& = (2\pi)^{\frac{N}{2}} \prod\limits_{i=0}^{N-1} \frac{\Gamma \left(1+(i+1)/\alpha\right)}{\Gamma \left(1+1/\alpha\right)}.
\end{split}
\end{equation}
The multivariate Hermite polynomials admit the expansion
\begin{equation}
\label{mult_H_exp}
H_\lambda^{\alpha}(x) =\sum_{\mu \subseteq \lambda }\omega_{\lambda \mu}^{\alpha} \, 
\frac{C_\mu^{(\alpha)}(x)}{C_\mu^{(\alpha)}\left(1^N\right)},
\end{equation}
where the coefficients $\omega^\alpha_{\lambda \mu}$ satisfy
\begin{equation}
\label{rec_H_c}
\begin{split}
\omega^{\alpha}_{\lambda \sigma} &= \frac{1}{\abs{\lambda}-\abs{\sigma}}\left(\sum_i \binom{\sigma^{(i)(i)}}{\sigma^{(i)}}
 \binom{\sigma^{(i)}}{\sigma} \omega^{\alpha}_{\lambda \sigma^{(i)(i)}}\right.\\
 & \quad + \left. \sum_{i <j}\left(\sigma_i -\sigma_j -\frac{1}{\alpha}\left(i-j\right)\right)\binom{\sigma^{(i)(j)}}{\sigma^{(j)}}\binom{\sigma^{(j)}}{\sigma}\omega^{\alpha}_{\lambda \sigma^{(i)(j)}}\right),
\end{split}
\end{equation}
where the inequality $i <j$ takes only admissible values.  It can be proved that this recurrence relation is well defined~\cite[Th.~2.25]{DES07}.  The normalization that leads to the formulae in Sec.~\ref{sec:gauss_mom} is given by the choice $\omega^\alpha_{\lambda \lambda}=C_\lambda^{(\alpha)}\left(1^N\right)$.   The recurrence~\eqref{rec_H_c} can be implemented numerically.  It is, however, quite difficult to manipulate~\eqref{rec_H_c} analytically and hence to extract information on the polynomials $H^\alpha_\lambda$ from  the behaviour of the coefficients $\omega^{\alpha}_{\lambda \mu}$.  Indeed to our knowledge an explicit formula for $H_\lambda^\alpha(0)$ does not exist except for $N=1$.

\section{A multivariate Hermite polynomial identity (by B.~Winn)}
\label{brian_proof}
In this appendix we give a proof of the identity \eqref{H_conj}:
\begin{equation}
\label{H_conj_Brian}
H^{\alpha}_{(1^k)}(0) = 
\begin{cases}
\frac{\alpha^{k/2}k!(k-1)!!}{(\alpha)_k}\binom{N}{k} & \text{if $k$ is even,}\\
0 & \text{if $k$ is odd.}
\end{cases}
\end{equation}

  In light of \eqref{av_jack_Hermite} it suffices to study the average
  \begin{equation}
    \label{eq:2}
    \frac{1}{\mathcal{N}^{(\mathrm{G})}_\alpha} \int_{\mathbb{R}^N} 
C_\lambda^{(\alpha)}(x)\prod_{j=1}^N e^{-\frac{x_j^2}{2}}
\prod_{1 \le j < k \le N}\abs{x_k - x_j}^{2/\alpha}d^N x. 
  \end{equation}
Our proof uses two main ingredients.  The first is a result of
Okounkov \cite{oko:poa} (which was conjectured by Goulden and
Jackson \cite[Conjecture 3.4]{GJ97}), which states that the value of
the integral in \eqref{eq:2} is
\begin{equation}
  \label{eq:1}
  \frac{j_\lambda}{\alpha^{3|\lambda|/2}|\lambda|!}C_\lambda^{(\alpha)}(1^N)
b(\lambda,\alpha),
\end{equation}
where $b(\lambda,\alpha)$ is the coefficient of $p_2(x)^{|\lambda|/2}$ in
the expansion of $C_\lambda^{(\alpha)}(x)$ if $|\lambda|$ is even, and 
$0$ otherwise. Thus, by \eqref{av_jack_Hermite},
\begin{equation}
  \label{eq:3}
  H_\lambda^\alpha(0) =   \frac{(-1)^{|\lambda|/2}j_\lambda}
{\alpha^{3|\lambda|/2}|\lambda|!} C_\lambda^{(\alpha)}(1^N) b(\lambda,\alpha).
\end{equation}
We can evaluate \eqref{eq:3} for $\lambda=(1^k)$, since by \eqref{eq:4}
\begin{equation}
  \label{eq:6}
  C_{(1^k)}^{(\alpha)} = \frac{\alpha^k k!}{c'((1^k),\alpha)}e_k,
\end{equation}
where $e_k$ is the $k$th elementary symmetric polynomial.  To find the value
of $b((1^k),\alpha)$ we use our second ingredient which is the expansion
of $e_k$ in the power-sum symmetric polynomial basis, which may be found
in e.g. \cite[Prop.~7.7.6]{sta:ecII}:
\begin{equation}
  \label{eq:7}
  e_k = \sum_{\mu\,
\vdash k} (-1)^{k-\ell(\mu)}z_\mu^{-1} p_\mu,
\end{equation}
where the quantity $z_\mu$ was defined immediately after
\eqref{scalar_product}.  We need the coefficient where $\mu=(2^{k/2})$ in
the expansion \eqref{eq:7} whence $\ell(\mu)=k/2$ and
\begin{equation}
  \label{eq:8}
  z_\mu = 2^{k/2} \left( \frac{k}2 \right)!.
\end{equation}
Then, provided $k$ is even,
\begin{equation}
  \label{eq:9}
  b((1^k),\alpha) = \frac{\alpha^k k!}{c'((1^k),\alpha)} \frac{(-1)^{k/2}}
{2^{k/2}(k/2)!}.
\end{equation}
The value of $C_{(1^k)}^{(\alpha)}(1^N)$ is given in \eqref{eq:5}, and
so putting together \eqref{eq:3}, \eqref{eq:5}, \eqref{j_const} and
\eqref{eq:9} we get
\begin{align}
  \label{eq:11}
    H_{(1^k)}^\alpha(0) &=   \frac{(-1)^{k/2}j_{(1^k)}}
{\alpha^{3k/2} k!} \frac{\alpha^{k} k!}{(\alpha)_k} \binom{N}{k}
 \frac{\alpha^k k!}{c'((1^k),\alpha)} \frac{(-1)^{k/2}}
{2^{k/2}(k/2)!} \nonumber \\
&= \frac{c((1^k),\alpha)\alpha^{k/2} k!}{(\alpha)_k 2^{k/2} (k/2)!} \binom{N}k.
\end{align}
From \eqref{cp_const} we have
\begin{equation}
  \label{eq:10}
  c((1^k),\alpha) = \prod_{s=1}^k (k-s+1) = k!,
\end{equation}
so that we end up with
\begin{equation}
  \label{eq:12}
    H_{(1^k)}^\alpha(0) =   
 \frac{\alpha^{k/2} (k!)^2 }{(\alpha)_k 2^{k/2} (k/2)!} \binom{N}{k},
\end{equation}
for $k$ even.  To finish note that
\begin{equation}
  \label{eq:13}
  \frac{k!}{2^{k/2}(k/2)!} = (k-1)!!
\end{equation}
for $k$ even.

\end{appendices}



 \bibliographystyle{amsplain}
\providecommand{\bysame}{\leavevmode\hbox to3em{\hrulefill}\thinspace}
\providecommand{\MR}{\relax\ifhmode\unskip\space\fi MR }
\providecommand{\MRhref}[2]{%
  \href{http://www.ams.org/mathscinet-getitem?mr=#1}{#2}
}
\providecommand{\href}[2]{#2}

\end{document}